\documentclass[ reprint,nofootinbib,amsmath,amssymb,onecolumn,aps]{revtex4-2}
\usepackage{appendix}
\usepackage{graphicx}
\usepackage{dcolumn}
\usepackage[colorlinks,linkcolor=magenta,anchorcolor=cyan,citecolor=blue]{hyperref}
\usepackage{bm}

\def\be{\begin{equation}}
\def\ee{\end{equation}}
\def\ba{\begin{eqnarray}}
\def\ea{\end{eqnarray}}

\def\dbar{{\mathchar '26\mkern -10mu\delta}}

   \def\y {\psi}              \def\x {\xi}        
              \def\.{\cdot}

\begin{document}
\title{Black hole thermodynamics is around the corner}
\author{Gerui Chen$^{1,2}$}
\email{t20152277@csuft.edu.cn}
\author{Wei Guo$^{2,3}$}
\email{guow@mail.bnu.edu.cn}
\author{Xin Lan$^{2,3}$}
\email{xinlan@mail.bnu.edu.cn}
\author{Hongbao Zhang$^{2,3}$}
\email{hongbaozhang@bnu.edu.cn}
\author{Wei Zhang$^{2,3}$}
\email{w.zhang@mail.bnu.edu.cn}
\affiliation{$^1$ College of Electronic Information and Physics, Central South University of Forestry and Technology, Changsha 410004, China\\
$^2$School of Physics and Astronomy, Beijing Normal University, Beijing 100875, China\\
$^3$ Key Laboratory of Multiscale Spin Physics, Ministry of Education, Beijing Normal University, Beijing 100875, China\\}

\date{\today}

\begin{abstract}
We propose to work on the Euclidean black hole solution with a corner rather than with the prevalent conical singularity. As a result, we find that the Wald formula for black hole entropy can be readily obtained for generic $F(R_{abcd})$ gravity by using both the action without the corner term and the action with the corner term due to their equivalence to the first order variation.
%which implies that it is the corner rather than the corner term that encodes the entropy related information. 
With such an equivalence, we further make use of a special diffeomorphism to accomplish a direct derivation of the ADM Hamiltonian conjugate to the Killing vector field normal to the horizon in the Lorentz signature as a conjugate variable of the inverse  temperature in the grand canonical ensemble.

%By working with the covariant phase space formalism, we have shown that not only can the Hamiltonian conjugate to a Killing vector field $\xi$ be expressed as the sum of the associated Noether charge and $\xi$ contracted with the Hilbert action boundary term for $F(R_{abcd})$ gravity, but also be written as its contraction with another $\xi$ independent tensor field. With this, we have proven the equivalence of Noether charge and Hilbert action boundary term formulae for the stationary black hole entropy in $F(R_{abcd})$ gravity, which is further substantiated by our explicit computation using both formulae. 
\end{abstract}
\maketitle
%\onecolumngrid

\section{Introduction}
 The four laws of black hole mechanics \cite{Bardeen}, due to Hawking's seminal discovery that black holes radiate thermally with the temperature proportional to the surface gravity of black hole horizon \cite{Hawking} as well as Bekenstein's original proposal that black holes should be  assigned an entropy proportional to the area of black hole horizon \cite{Bekenstein}, is promoted as the four laws of black hole thermodynamics, which not only provides some of the deepest insights into the fundamental nature of black holes, but also offers us a unique key to the formulation of quantum theory of gravity. 
 
 Among others, Euclidean approach to quantum gravity was proposed \cite{Hartle,GH}. In particular, with the success in reproducing the Bekenstein-Hawking entropy for the black hole in general relativity, Euclidean approach to black hole thermodynamics demonstrates its remarkable power in disclosing the thermal nature of black holes \cite{HP,HH,GPP}. However, as commented most recently in \cite{Witten}, the original derivation is quite a coup, where only at the end of the computation can one see the black hole entropy proportional to the horizon area. Such a deficiency was rescued by the so-called conical deficit angle method \cite{CT,BTZ,SU}, whereby the resulting black hole entropy proportional to the horizon area becomes manifest due to the fact the scalar curvature develops a conical singularity with a $\delta$ function supported on the horizon. But as criticized in \cite{Iyer}, the resulting action is mathematically ill defined for the generic gravity theory with the Lagrangian form beyond the linear order in curvature.
Later on, it was shown in \cite{FS} that the conical deficit angle method could also be applied to calculate the black hole entropy in the more general gravity theory somehow by viewing the conical singularity as the appropriate limit of the converging sequences of regular spaces.
%However, %not only is the involved computation coordinate dependent, but also extraordinarily tedious. Moreover,
 %In this sense, introducing the conical singularity amounts to opening Pandora's box. So 
 But nevertheless, it is still highly desirable to develop an alternative method to such a conical deficit angle one. 

With this in mind, we propose to work on the Euclidean black hole solution with a corner rather than with the conical singularity. As a result, neither singularity is created nor regularization is needed in our recipe, whereby not only can we reproduce the black hole entropy for generic $F(R_{abcd})$ gravity in the grand canonical ensemble, but also provide a direct derivation of the ADM (Arnowitt-Deser-Misner) Hamiltonian conjugate to the Killing vector field normal to the black hole horizon as a conjugate variable to the inverse of black hole temperature for black hole thermodynamics in the grand canonical ensemble. In hindsight, although the conical deficit angle method captures the essential role played by the black hole horizon somehow, we do not think that it is the simplest machinery, if not mathematically ill defined. Instead, we think that the manifold with a corner is the right setup to work on for Euclidean approach to black hole thermodynamics.

The structure of this paper is organized as follows. In the subsequent section, we shall present the generic structure for the variation of Lagrangian form on the Euclidean manifold with a corner, whereby we introduce one action without the corner term and the other action with the corner term as well as their variations on top of the solution space. Then in Section \ref{derivation}, we reproduce the Wald entropy formula for black holes using both actions mentioned above in the grand canonical ensemble, because the two actions are equal to each other to the first order variation. Furthermore, by pulling back the action with the corner term using a special diffeomorphism, we achieve a direct derivation of the ADM Hamiltonian conjugate to the Killing vector field normal to the black hole horizon as a conjugate variable to the inverse temperature. We conclude our paper with some discussions in the last section. 

We will follow the notation and conventions of \cite{GR}. In addition, we shall use the boldface letters to denote differential forms with the tensor indices suppressed. 

\section{Variation of the Lagrangian form on the manifold with a corner}

Let us start from the generic $F(R_{abcd})$ gravity with the Lagrangian form given by 
\begin{equation}
\mathbf{L}=\bm{\epsilon}F(R_{abcd},g_{ab}),
\end{equation}
 where $\bm{\epsilon}$ is the spacetime volume and $F$ is  an arbitrary function of the Riemann tensor $R_{abcd}$ and the metric $g_{ab}$. Its variation reads 
\begin{equation}\label{variationradius}
    \delta \mathbf{L}= \bm\epsilon E_g^{ab}\delta g_{ab}+d\mathbf{\Theta}.
\end{equation}
 Here
\begin{eqnarray}
E_g^{ab}=\frac{1}{2}g^{ab}F+\frac{1}{2}\frac{\partial F}{\partial g_{ab}}+2\nabla_{c}\nabla_{d}\psi^{c(ab)d}
\end{eqnarray}
with $E_g^{ab}=0$ corresponding to the equation of motion,
and $\mathbf{\Theta}=\vartheta\cdot\bm{\epsilon}$ is the symplectic potential with 
 \begin{equation}
\vartheta^a=2(\nabla_d\psi^{bdca}\delta g_{bc}-\psi^{bdca}\nabla_d\delta g_{bc}),
\end{equation}
where the dot denotes the contraction of a vector with the first index of the differential form, and $\psi^{abcd}$ is defined as the derivative of $F$ with respect to $R_{abcd}$ by pretending that it is independent of the metric, namely
$\psi^{abcd}\equiv \frac{\partial F}{\partial R_{abcd}}$ \cite{Zhang0,Zhang1,HB,Zhang2}.

\begin{figure}
\centering
%\begin{minipage}{0.02\textwidth}
  % \end{minipage}
\includegraphics[width=0.5\textwidth]{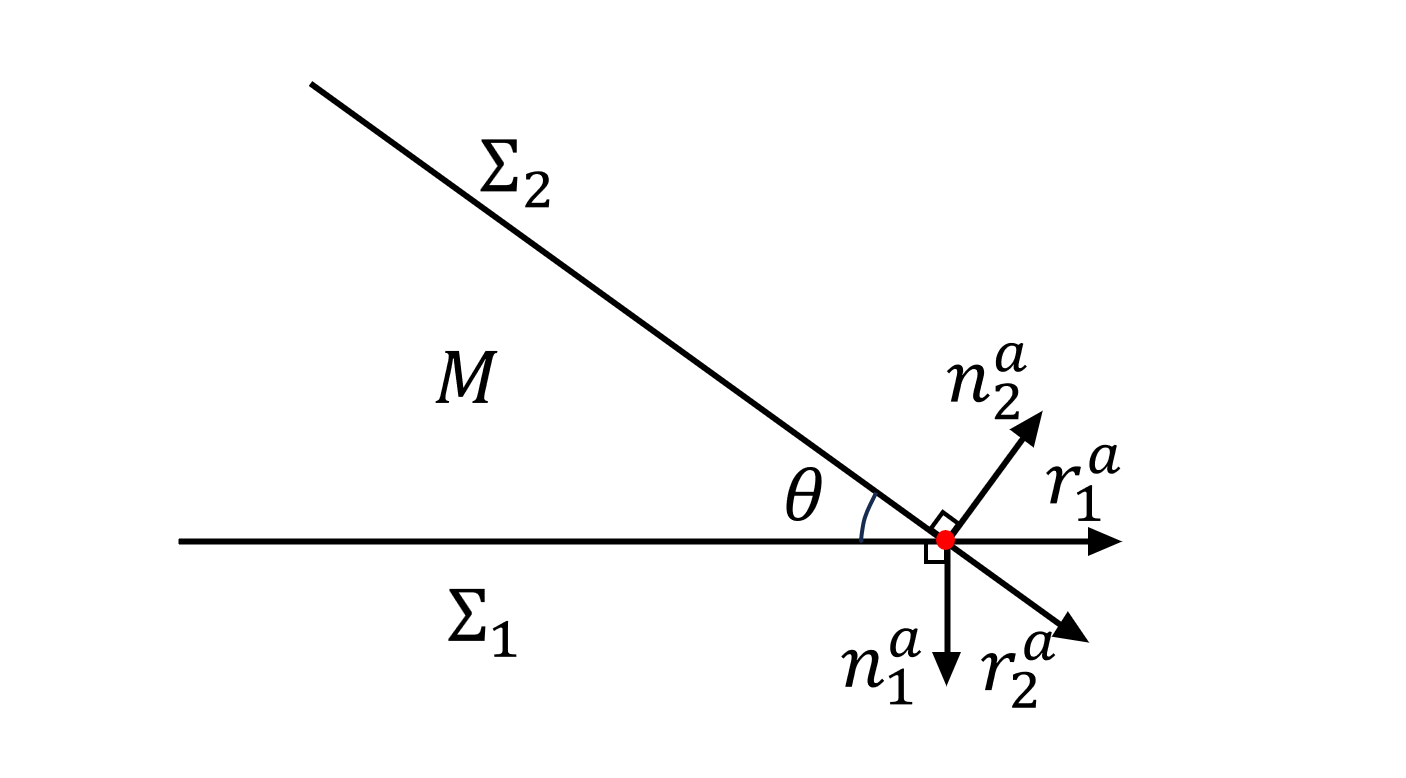}
\caption{The portion of the boundary of the space $M$ is given by $\Sigma_1$ and $\Sigma_2$, which intersect with each other at a co-dimension $2$ corner denoted by the red point. $n_i^a$  and $r_i^a$ with $n_i^a r_{ia}=0$ are transverse orthonormal vectors at the corner, where $n_i^a$ is the normal vector to $\Sigma_i$ with $i=1,2$.}\label{p}
\end{figure}

As illustrated in Fig. \ref{p}, now let us consider a Euclidean manifold $M$ with the portion of its boundary given by $\Sigma_1$
and $\Sigma_2$. $\Sigma_1$ and $\Sigma_2$ are further assumed to intersect with each other at the corner $\mathcal{S}$. With the outward-pointing unit normal vector and the induced metric of $\Sigma_i$ denoted respectively as $n_{ia}$ and $h_{iab}$, the variation of the metric can be expressed as 
\begin{equation}\label{ex1}
    \delta g^{ab}|_{\Sigma_i}=-2\delta a_in_i^an_i^b+\dbar{A}_i^an_i^b+\dbar{A}_i^bn_i^a+\delta h_i^{ab}
\end{equation}
with $n_{ia}\dbar{A}_i^a=0$, where we have worked in the gauge in which $\Sigma_i$ remain fixed under variation such that $\delta n_{ia}=\delta a_in_{ia}$\footnote{Here $\dbar$ denotes the incomplete variation with $\dbar A_i^a$ differing from those appearing in \cite{Zhang0,Zhang1,HB,Zhang2} by a minus sign. In addition, readers are suggested to refer to the Appendix for the detailed derivation of Eqs. (\ref{ex1}), (\ref{ex0}), 
 (\ref{re2}) and (\ref{corner}) if interested.}. In particular, at the corner we have 
%the metric can be written as follows
%\begin{equation}
%g_{ab}|_\mathcal{S}=\csc^2\theta(n_{1a}n_{1b}+n_{2a}n_{2b})+\cos\theta\csc^2\theta(n_{1a}n_{2b}+n_{2a}n_{1b})+\sigma_{ab},
%\end{equation}
 %Furthermore, one can show that 
\begin{eqnarray}\label{ex0}
&&\delta h_1^{ab}|_\mathcal{S}=2(\cot\theta\delta\theta-\delta a_2)r_1^ar_1^b+r_1^a\dbar\tilde{B}_1^b+\dbar\tilde{B}_1^ar_1^b+\delta \gamma^{ab},\nonumber\\
&&\delta h_2^{ab}|_\mathcal{S}=2(\cot\theta\delta\theta-\delta a_1)r_2^ar_2^b+r_2^a\dbar\tilde{B}_2^b+\dbar\tilde{B}_2^ar_2^b+\delta \gamma^{ab},
\nonumber\\
&&\dbar{A}_1^a|_\mathcal{S}=[\delta\theta+\cot\theta(\delta a_2-\delta a_1)]r_1^a+\dbar\tilde{A}_1^a,\nonumber\\
&&\dbar{A}_2^a|_\mathcal{S}=[\delta\theta+\cot\theta(\delta a_1-\delta a_2)]r_2^a+\dbar\tilde{A}_2^a,
\end{eqnarray}
where $\theta$ is the subtended angle of our manifold $M$ at the corner with $\cos\theta=-n_1^a n_{2a}$, $\gamma^{ab}$ is the induced metric on it, $r_i^a$ with $r_i^an_{ia}=0$ are outward pointing unit vectors normal to the corner on $\Sigma_i$, satisfying
\begin{equation}\label{re1}
\begin{pmatrix}
n_2^a \\
r_2^a
\end{pmatrix}
=
\begin{pmatrix}
-\cos\theta & \sin\theta \\
\sin\theta & \cos\theta
\end{pmatrix}
\begin{pmatrix}
n_1^a \\
r_1^a
\end{pmatrix},\,
\begin{pmatrix}
n_1^a \\
r_1^a
\end{pmatrix}
=
\begin{pmatrix}
-\cos\theta & \sin\theta \\
\sin\theta & \cos\theta
\end{pmatrix}
\begin{pmatrix}
n_2^a \\
r_2^a
\end{pmatrix},
\end{equation}
%\begin{equation}
 %   n_1^a=\sin\theta r_2^a-\cos\theta n_2^a,\quad r_1^a=\cos\theta r_2^a+\sin\theta n_2^a,
%\end{equation}
and $\dbar\tilde{A}_i^a$ as well as $\dbar\tilde{B}_i^a$ are tangential to the corner, satisfying the following relation
\begin{equation}\label{re2}
\dbar{\tilde B_1^a}=\cot\theta\dbar{\tilde A_1^a }+\csc\theta\dbar{\tilde A_2^a},\quad 
\dbar{\tilde  B_2^a}=\csc\theta\dbar{\tilde A_1^a}+\cot\theta\dbar{\tilde A_2^a }.
\end{equation}

On the boundary $\Sigma_i$, $\mathbf{\Theta}$ can be cast into the following form \cite{Zhang0,Zhang1,HB,Zhang2}
\begin{eqnarray}\label{boundaryex}
    \mathbf{\Theta}|_{\Sigma_i}=-\delta \mathbf{B}+d\mathbf{C}+\mathbf{F},
\end{eqnarray}
where
\begin{eqnarray}\label{onlyc}
    \mathbf{B}=4\Psi_{ab}K^{ab}\hat{\bm\epsilon},\quad
    \mathbf{C}=\mathbf\omega\cdot\hat{\bm\epsilon},\quad
    \mathbf{F}=\hat{\bm\epsilon}(T_{hbc}\delta h^{bc}+T_{\Psi bc}\delta\Psi^{bc})
\end{eqnarray}
with $K_{ab}=h_a{}^c\nabla_cn_b$ the extrinsic curvature, $\hat{\bm\epsilon}$ the induced volume defined as  $\bm\epsilon=\mathbf n\wedge \hat{\bm\epsilon}$, $\Psi_{ab}=\psi_{acbd}n^cn^d$, and
\begin{eqnarray}\label{bc}
    \omega^a&=&2\Psi^a{}_b\dbar A^b+2h^{ae}\psi_{ecdb}n^d\delta h^{bc},\nonumber\\
    T_{hbc}&=&-2\Psi_{de}K^{de}h_{bc}+2n^a\nabla^e\psi_{deaf}h^d{}_{(b}h^f{}_{c)}-2 \Psi_{a(b}K^a{}_{c)}-2D^a(h_a{} ^eh_{(c}{}^f\psi_{|efd|b)}n^d),\nonumber\\
    T_{\Psi bc}&=&4 K_{bc}.
\end{eqnarray}
Here for the notational convenience, we have suppressed the index $i$. It is supposed to be instructive for us to pause a little bit to extract the physical meaning of $\mathbf{B}$, $\mathbf{C}$, and $\mathbf{F}$ appearing in Eq. (\ref{boundaryex}) by taking Einstein's gravity as an example. It follows from the Lagrangian form $\mathbf{L}=-\frac{1}{16\pi}R\bm\epsilon$ that $\psi_{abcd}=-\frac{1}{32\pi}(g_{ac}g_{bd}-g_{ad}g_{bc})$ and $\Psi_{ab}=-\frac{1}{32\pi}h_{ab}$, whereby we have
\begin{equation}
    \mathbf{B}=-\frac{K}{8\pi}\hat{\bm\epsilon},\quad \mathbf{C}=-\frac{1}{16\pi}\dbar A\cdot\hat{\bm\epsilon},\quad \mathbf{F}=\frac{1}{2}T_{bc}\delta h^{bc}\hat{\bm\epsilon}
\end{equation}
with $T_{bc}=-\frac{1}{8\pi}(K_{bc}-Kh_{bc})$ the familiar Brown-York boundary energy momentum tensor. Whence we know that $\mathbf{B}$ gives rise to the well-known Gibbons-Hawking-York (GHY) boundary term and $\mathbf{F}$ is related to the Brown-York boundary energy momentum tensor. $\mathbf{C}$ may be a little bit unfamiliar but will play an important role in our later discussion. We like to have a better understanding of this term as follows. 
By Stokes theorem, the $d\mathbf{C}$ terms from both $\Sigma_1$ and $\Sigma_2$ contribute to the corner. Then by a straightforward calculation with the aid of Eq. (\ref{re1}) and Eq. (\ref{re2}) as well as $\psi_{abcd}n_1^ar_1^b=-\psi_{abcd}n_2^ar_2^b$, we obtain
\begin{equation}\label{corner}
    \mathbf{C}|_\mathcal{S}=\mathbf{C}_1+\mathbf{C}_2=\delta\theta\psi^{abcd}\varepsilon_{ab}\varepsilon_{cd} \tilde{\bm\epsilon}+2\delta\gamma^{bc}(r_1^an_1^d+r_2^an_2^d)\psi_{acdb}\tilde{\bm\epsilon},
\end{equation}
where $\varepsilon_{ab}=(n\wedge r)_{ab}$ is the binormal and $\tilde{\bm\epsilon}$ is the induced volume at the corner with $\bm{\epsilon}=\bm\varepsilon\wedge\tilde{\bm{\epsilon}}$\footnote{Although the binormal from $\Sigma_1$ and its induced volume at the corner differ from those from $\Sigma_2$ by a minus sign, Eq. (\ref{corner}) holds no matter which binormal and induced volume are chosen.} . For Einstein's gravity, the second term vanishes automatically due to $r_i^an_{ia}=0$ as well as $\delta\gamma^{bc}r_{ib}n_{ic}=0$, and we have 
\begin{equation}\label{suppl}
    \mathbf{C}|_\mathcal{S}=-\frac{1}{8\pi}\delta\theta\tilde{\bm{\epsilon}}.
\end{equation}
It is noteworthy that for a generic $F(R_{abcd})$ gravity, the second term in Eq. (\ref{corner}) does not vanish.
%Note that the second term in the above equation does not vanish generically. But it vanishes automatically for general relativity, because the Einstein-Hilbert action gives rise to $\psi_{acdb}\propto(g_{ad}g_{cb}-g_{ab}g_{cd})$. 
But it vanishes when evaluated on the background with the subtended angle equal to $2\pi$ because of $n_1^a=-n_2^a$ and $r_1^a=r_2^a$ or with the subtended angle equal to $\pi$ because of $n_1^a=n_2^a$ and $r_1^a=-r_2^a$\footnote{When the corner has the subtended angle equal to $\pi$ or $2\pi$, the corresponding Euclidean manifold can also be regarded as  a regular manifold without a corner. So if one varies this Euclidean manifold within such a class of manifolds without a corner, then the first term in Eq. (\ref{corner}) also vanishes as it should be the case.}. 

With the above preparation, we now supplement the bulk action with the generalized GHY boundary term $\mathbf{B}$ as follows
\begin{equation}
    I=\int_M\mathbf{L}+\int_{\partial M}\mathbf{B}.
\end{equation}
Then it follows from Eq. (\ref{variationradius}) and Eq. (\ref{boundaryex}) that the variation of the action $I$ on top of the solution space reads
\begin{equation}\label{firstone}
    \delta I=\int_{\partial M}\mathbf{F}+\int_\mathcal{S}\mathbf{C},
\end{equation}
where the last term should be understood as the contribution from all the corners. Among others, Eq. (\ref{firstone}) tells us that the conjugate variable to the subtended angle is given by $\int_\mathcal{S}\psi^{abcd}\varepsilon_{ab}\varepsilon_{cd}\tilde{\bm\epsilon}$. As such,  we like to  make a Legendre transformation by supplementing the above action with the additional corner term as follows
\begin{equation}\label{ok}
    I'=I+I_\mathcal{S}
\end{equation}
with 
\begin{equation}
    I_\mathcal{S}=(\theta_0-\theta)\int_\mathcal{S}\psi^{abcd}\varepsilon_{ab}\varepsilon_{cd}\tilde{\bm\epsilon},
\end{equation}
where the integral constant $\theta_0$ will be chosen as the subtended angle of the corners appearing in the solution space away from which we make variation. So when evaluated on such a solution space, not only $I_\mathcal{S}=0$, namely $I'=I$, but also the variation of $I'$ reads
\begin{eqnarray}\label{anotherguy}
    \delta I'&=&\int_{\partial M}\mathbf{F}+(\theta_0-\theta)\int_\mathcal{S}\delta(\psi^{abcd}\varepsilon_{ab}\varepsilon_{cd} \tilde{\bm\epsilon})+2\int_\mathcal{S}\delta\gamma^{bc}(r_1^an_1^d+r_2^an_2^d)\psi_{acdb}\tilde{\bm\epsilon}\nonumber\\
    &=&\int_{\partial M}\mathbf{F}+2\int_\mathcal{S}\delta\gamma^{bc}(r_1^an_1^d+r_2^an_2^d)\psi_{acdb}\tilde{\bm\epsilon},
\end{eqnarray}
%on top of the solution space. Note that in what follows there is essentially no corner contribution to the above action for the solution space on top of which we make the variation, so the natural choice of the integral constant $\theta_0$ is the subtended angle of the corner appearing over there. Accordingly, the above equation reduces to 
%\begin{equation}\label{anotherguy}
 %   \delta I'=,
%\end{equation}
which vanishes when the Dirichlet boundary condition is imposed on the boundary as well as at the corner. In this sense, the primed action $I'$ with the additional corner term satisfies the variational principle, which is virtually the underlying motivation for the introduction of this corner term in Einstein's gravity \cite{Hayward}  as well as the proposal to calculate the generalized gravitational entropy \cite{LM} for fixed area states by sticking to $I'$ \cite{TT,BMZ,ABM,KS1,KS2}. On the other hand, the unprimed action $I$ has also been used to calculate the generalized gravitational entropy for Hartle-Hawking states in \cite{KS1,KS2}.  In the next section, we shall show that not only can the black hole entropy be readily obtained by working with $I'$, but also by working with $I$. Moreover, the reason why one can resort to $I'$ comes essentially from the fact that $I'$ is equal to $I$ to the order we are interested in.
%We suspect that such an argument can be applied similarly to the calculation of the generalized gravitational entropy, which is supposed to have nothing to do with the aforementioned variational principle claimed in the previous literature.

\section{Derivation of black hole entropy and ADM Hamiltonian}\label{derivation}
According to the Euclidean approach to black hole thermodynamics,
the Gibbs free energy of the black hole is given by \cite{HB}
\begin{equation}\label{sp}
    \beta G=[I(\beta)]
\end{equation}
with the background subtracted action $[I(\beta)]=I(\beta)-I^0(\beta)$, where $I(\beta)$ denotes the on shell action of the regular Euclidean black hole with the temperature $T=\frac{1}{\beta}$ and $I^0(\beta)$ denotes the corresponding quantity for the reference space. Thus the black hole 
entropy can be obtained as
\begin{equation}\label{thermo}
    S=(\beta\partial_\beta-1)[I(\beta)]|_{\beta_0}.
\end{equation}
It is noteworthy that this is precisely the original definition for the black hole entropy in the Euclidean approach to black hole thermodynamics, whereby Hawking and his companions wind up with the celebrated Bekenstein-Hawking formula for Einstein's gravity. However, such a derivation looks miraculous at first glance since the contribution to $I(\beta)$ comes only from the bulk as well as the asymptotical surface at infinity and has no direct bearing on the black hole horizon. To demystify this, we like to resort to the translation isometry along the imaginary time direction, whereby one can rewrite $I(\beta)=\frac{\beta}{\beta_0}I_{\beta_0}(\beta)$ with $I_{\beta_0}(\beta)$ the action evaluated on the Euclidean black hole of the inverse temperature $\beta$ with the imaginary time interval $[0,\beta_0]$. Accordingly, the above entropy formula reduces to 
\begin{equation}\label{trick}
    S=\beta\partial_\beta [I_{\beta_0}(\beta)]|_{\beta_0}.
\end{equation}
\begin{figure}
\centering
%\begin{minipage}{0.02\textwidth}
  % \end{minipage}
\includegraphics[width=0.9\textwidth]{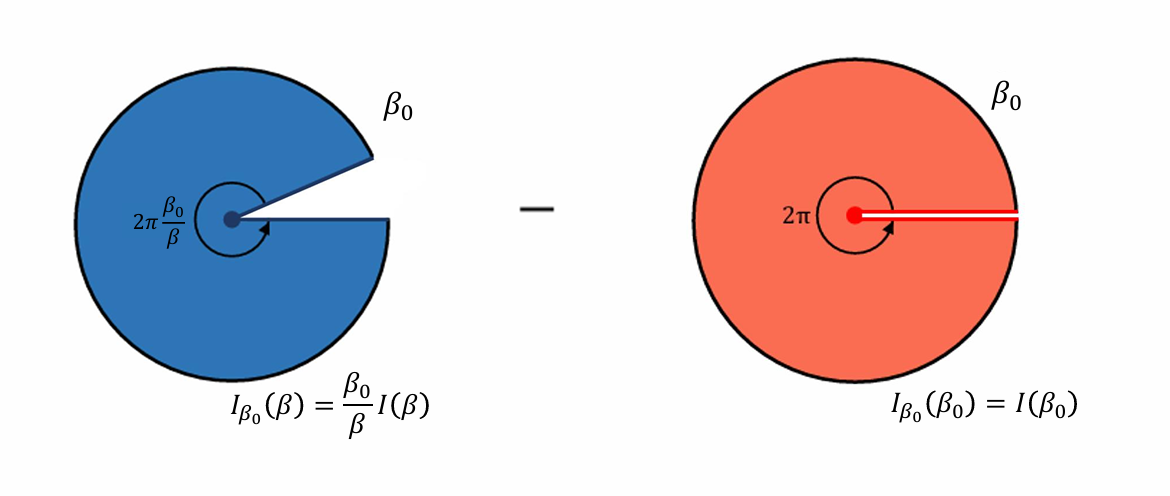}
\caption{The two Euclidean black holes share the same imaginary time interval $\beta_0$, where the black hole in blue has $T=\frac{1}{\beta}$ while the black hole in orange has $T_0=\frac{1}{\beta_0}$.
The boundary consists of two cuts and the asymptotical surface with three corners. The bifurcation surface is identified as the inner corner from the intersection of the two cuts and the two outer corners arise from the intersection of the asymptotical surface with each cut, respectively. }\label{p1}
\end{figure}

As demonstrated in Fig. \ref{p1}, $I_{\beta_0}(\beta)$ is for the  black hole solution with the subtended angle $\theta=2\pi\frac{\beta_0}{\beta}$ at the inner corner. Thus it follows from Eq. (\ref{firstone}) that at the first order of $(\beta-\beta_0)$
\begin{equation}\label{algebra}
   [I_{\beta_0}(\beta)]-[I_{\beta_0}(\beta_0)]=2\pi(\frac{\beta_0}{\beta}-1)\int_\mathcal{B}\psi^{abcd}\varepsilon_{ab}\varepsilon_{cd}\tilde{\bm\epsilon},
\end{equation}
with $\mathcal{B}$ the bifurcation surface, where we have used the simple observation that not only do the contributions from the two cuts cancel out each other but also the contributions from the two outer corners cancel out each other as well as the important fact that the contribution from the asymptotical surface vanishes no matter whether the asymptotical geometry is flat or Anti-de Sitter (AdS)\footnote{Note that there is neither cut nor corner for the reference space, so the contribution comes solely from the asymptotical surface.} \cite{HB}. Then Eq. (\ref{trick}) gives rise to the black hole entropy as
\begin{equation}
S=-2\pi\int_\mathcal{B}\psi^{abcd}\varepsilon_{ab}\varepsilon_{cd}\tilde{\bm\epsilon}.
\end{equation}
According to the familiar relationship $\tau=it$ as well as $\mathbf{L}=-i\mathbf{L}_L$ between the Euclidean space and Lorentzian space, the above result is exactly equivalent to the Wald formula for the black hole entropy, derived originally in the Lorentz signature based on the first law of black hole thermodynamics \cite{Wald,IW}.

\begin{figure}
\centering
%\begin{minipage}{0.02\textwidth}
  % \end{minipage}
\includegraphics[width=0.4\textwidth]{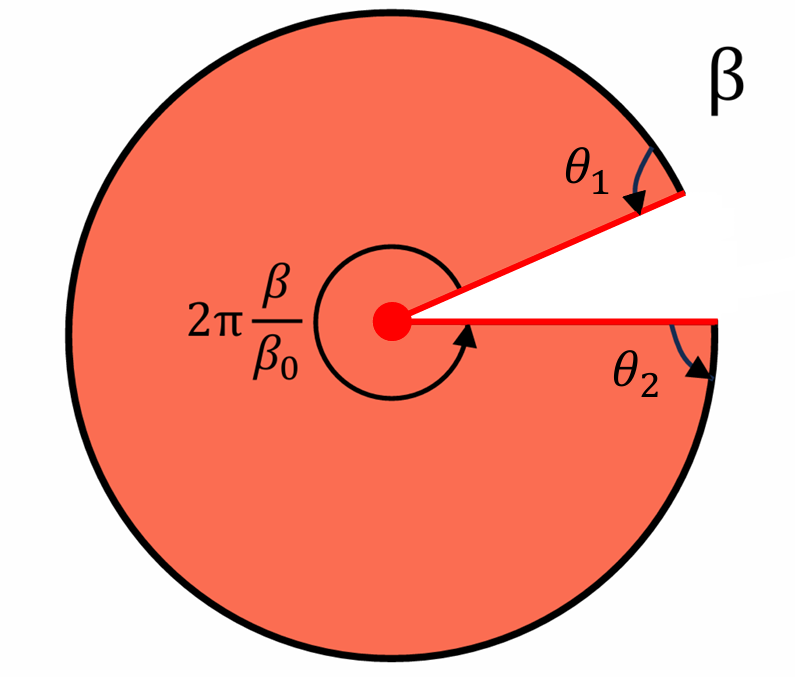}
\caption{The Euclidean black hole at the temperature $T_0=\frac{1}{\beta_0}$ but with the imaginary time interval $\beta$. }\label{p2}
\end{figure}

On the other hand, note that Eq. (\ref{algebra}) can be rewritten as 
\begin{equation}
   [I(\beta)]=\frac{\beta}{\beta_0}[I(\beta_0)]+2\pi(1-\frac{\beta}{\beta_0})\int_\mathcal{B}\psi^{abcd}\varepsilon_{ab}\varepsilon_{cd}\tilde{\bm\epsilon}=[I_\beta(\beta_0)]+(2\pi-2\pi\frac{\beta}{\beta_0})\int_\mathcal{B}\psi^{abcd}\varepsilon_{ab}\varepsilon_{cd}\tilde{\bm\epsilon}
\end{equation}
to the first order of $(\beta-\beta_0)$, which well explains why one can also derive the black hole entropy using Eq. (\ref{thermo}) with $I(\beta)$ replaced by $I'_\beta(\beta_0)$, namely the action with the corner term Eq. (\ref{ok}) evaluated on the black hole of $T_0=\frac{1}{\beta_0}$ but with the imaginary time interval $\beta$ along the imaginary time $\tau$, as depicted in Fig. \ref{p2}. Here as pointed out at the end of the previous section, not only have we taken the integral constant $\theta_0=2\pi$ at the inner corner, but also taken the subtended angle for each outer corner as $\theta_0$ over there such that there is no corner term from each outer corner. In addition, we would like to identify the similarity and difference between this alternative method with the corner term and the conical deficit angle method. The similarity lies in the fact that both of them depends solely on the black hole solution with $T_0$ rather than any other black hole solution with a different temperature. The difference is that the additional corner term here is obviously finite for an arbitrary $F(R_{abcd})$ gravity theory while the additional contribution arising from the conical singularity at the corner is divergent and needs regularization except for Einstein's gravity. 

It is noteworthy that we can also derive the conjugate variable to the inverse temperature $\beta$ directly from $I'_\beta(\beta_0)$. To this end, we like to consider the diffeomorphism $\phi_\beta$ generated by $\tau'=\tau+(\beta-\beta_0)f(\tau)$, where $f(\tau)$ with $\tau\in[0,\beta_0]$ is a smooth function of the imaginary time, exactly equal to zero in the neighborhood of $\tau=0$ and equal to one in the neighborhood of $\tau=\beta_0$\footnote{Note that such a diffeomorphism is not allowed in the Euclidean manifold with the conical singularity because it requires the infinitesimal generator $\xi^a$ at $\tau=0$ and $\tau=\beta_0$ be exactly the same.}. Such a design leads to two nice properties. One is that the corresponding diffeomorphism pulls the region $[0,\beta]$ back to the region $[0,\beta_0]$. The other is that not only is $\xi^a=f(\tau)(\frac{\partial}{\partial\tau})^a$ tangential to the asymptotical surface but also is a Killing vector field in both neighborhoods of $\tau=0$ and $\tau=\beta_0$. Accordingly, it follows from (\ref{anotherguy}) that 
\begin{equation}
    \partial_\beta I'_\beta(\beta_0)|_{\beta_0}=\int_\infty \hat{\bm\epsilon}(T_{hbc}\mathcal{L}_\xi h^{bc}+T_{\Psi bc}\mathcal{L}_\xi\Psi^{bc}),
\end{equation}
because all the other terms vanish automatically\footnote{The contributions from the two cuts and two outer corners vanish because there is no variation arising from our diffeomorphism over there due to the fact that $\xi^a$ is a Killing vector field therein. On the other hand, the contribution from the inner corner also vanishes because the variation evaluated on top of the regular Euclidean black hole solution with the subtended angle equal to $2\pi$.}. Then by the fact that $\mathcal{L}_\xi h^{bc}=-2D^{(b}\xi^{c)}$ and $\mathcal{L}_\xi\Psi^{bc}=\xi^dD_d\Psi^{bc}-\Psi^{dc}D_d\xi^b-\Psi^{bd}D_d\xi^c$ as well as the integration by parts, we further have 
\begin{equation}
\partial_\beta I'_\beta(\beta_0)|_{\beta_0}=\int_\infty [d(q_\xi \cdot \hat{\bm\epsilon})+\xi^f\Lambda_f\hat{\bm\epsilon}],
\end{equation}
where 
\begin{eqnarray}\label{brownyork}
    q_\xi^a=-(2T_h{}^a{}_c+2\Psi^{ab}T_{\Psi cb})\xi^c,\quad
    \Lambda_f=2D^bT_{hbf}+T_{\Psi bc}D_f\Psi^{bc}
+2D_d(T_{\Psi fc}\Psi^{dc}).
\end{eqnarray}
Note that 
$\Lambda_f$ evaluated on the solution space vanishes automatically \cite{Zhang1}, so we have
\begin{equation}
    \partial_\beta I'_\beta(\beta_0)|_{\beta_0}=\int_{\beta_0}q_\xi\cdot\hat{\bm\epsilon},
\end{equation}
where the orientation at the corner $\beta_0$ is determined by $\tilde{\bm\epsilon}$ with $\hat{\bm\epsilon}=d\tau\wedge\tilde{\bm{\epsilon}}$. According to the aforementioned relationship between the Euclidean space and Lorentzian space, one ends up with
\begin{equation}
    \partial_\beta I'_\beta(\beta_0)|_{\beta_0}=H_{\frac{\partial}{\partial t}}
\end{equation}
with $H_{\frac{\partial}{\partial t}}$ the ADM Hamiltonian conjugate to the Killing vector field $\frac{\partial}{\partial t}$ normal to the horizon of the black hole in the Lorentz signature \cite{Zhang1}. This result is consistent with $\partial_\beta[I(\beta)]=[H_{\frac{\partial}{\partial t}}]$ obtained from Eq. (\ref{sp}) as it should be the case. 

%But nevertheless, as evidenced by the final correct result it gives rise to,  the background subtraction method seems also applicable to the circumstances where such an embedding is impossible.

%Among others, one main purpose of this paper is to examine the applicability of the background subtraction method by working with the covariant phase space formalism. As a result, we succeed in identifying the necessary and sufficient condition for the validity of the background subtraction method. Then we further show that the resulting criterion is satisfied not only by Einstein's gravity but also by its higher derivative corrections for both black holes in asymptotically flat and AdS spacetimes, even though the induced metrics of the stationary black holes and the reference spacetime are not exactly the same on the boundary. With this in mind,  our result, on the one hand, offers a priori rather than a posteriori justification for the applicability of the background subtraction method in the aforementioned more generic circumstances. Accordingly, the background subtraction method is supposed to be as applicable as the covariant counterterm method. With this in mind, we further apply the background subtraction method to calculate the higher derivative corrections to Kerr-AdS black hole thermodynamics, where we find that the relevant computation can be simplified greatly by resorting to the spinor decomposition of Weyl tensor. 

\section{Conclusion}
Working on the Euclidean black hole solution with a corner rather than with a conical singularity, we have accomplished reproducing the Wald formula for the black hole entropy in an elegant manner, where the Euclidean action we use can be either the action without the corner term or the action with the corner term because they turn out to be equal to each other to the first order variation. %This observation tells us in a affirmative manner that it is the corner rather than the variational principle induced corner term that encodes the entropy related information, sharpening the prevalent perspective in the community.
By further resorting to the above equality, we are the first to achieve a direct derivation of the ADM Hamiltonian conjugate to the Killing vector field normal to the black hole horizon in the Lorentz signature as a conjugate variable to the inverse temperature by a special diffeomorphism, which otherwise would be not allowed in the Euclidean black hole solution with a conical singularity.

However, an arbitrary diffeomorphism covariant Lagrangian form of gravity also includes the derivatives of Riemann tensor. So it is desirable to see whether our recipe can also be extended to such a more general theory of gravity, where the key lies in whether one can also write $\bm\Theta$ explicitly in the form of Eq. (\ref{boundaryex}). In addition, so far we have restricted ourselves onto the on shell black hole solution with a corner. It is obviously important for one to explore the loop corrections to black hole thermodynamics by path integral of the fields propagating in the aforementioned on shell black hole solution with a corner, where the presence of the corner may display a universal effect on the loop corrections.  Another potential application of our recipe with a corner is to refine the derivation of the generalized gravitational entropy. Among others, there is a very limitation on the previous treatment \cite{LM,TT,BMZ,ABM,KS1,KS2},
in the sense that their boundary data at the asymptotical surface are assumed implicitly or explicitly to be the same, for instance for $I_{\beta_0}(\beta)$ and $I(\beta_0)$ such that there is no contribution to the variation of the action from the asymptotical surface automatically. Compared to this, we are not required to restrict ourselves into this scenario. Instead, by drawing on the most recent advance in the background subtraction method achieved in \cite{HB}, we can allow the ineludible difference between the boundary data for $I_{\beta_0}(\beta)$ and $I(\beta_0)$ but keep the zero contribution to the variation of the background subtracted action from the asymptotical surface. It is tempting to apply such a similar strategy to address the generalized gravitational entropy without the translation isometry along the imaginary time. 
%In addition, it is highly possible to achieve a Euclidean proof of the first law of generalized gravitational entropy by devising a special diffeomorphism. 

\begin{acknowledgments}
We are grateful to Edward Witten for valuable communications regarding his recent review paper on black hole thermodynamics, which spurs the initiation of this project. 
This work is partially supported by the National Key Research and Development Program of China with Grant No. 2021YFC2203001 as well as the National Natural Science Foundation of China with Grant Nos. 12361141825, 12475049, and 12575047.
%We like to thank Ran Li for his useful discussion on the conical singularity.

\end{acknowledgments}

%\newpage

%\onecolumngrid
\appendix
\section{Derivation of Eqs. (\ref{ex1}), (\ref{ex0}), 
 (\ref{re2}) and (\ref{corner})}\label{4D}

 First, $n_i^an_{ia}=1$ implies
\begin{equation}
    0=\delta (n_i^an_{ia})=\delta n_i^an_{ia}+\delta a_i,
\end{equation}
which gives rise to 
\begin{equation}
    \delta n_i^a=-\delta a_in_i^a+\dbar A_i^a
\end{equation}
with $\dbar A_i^a$ the tangential component of $\delta n_i^a$ to $\Sigma_i$. Then according to 
\begin{equation}
    g^{ab}|_{\Sigma_i}=n_i^an_i^b+h_i^{ab},
\end{equation}
we end up with Eq. (\ref{ex1}).

On the other hand, according to Eq. (\ref{re1}), we have 
$r_1^a n_{2a}=\sin\theta$. This implies
\begin{equation}
 \cos\theta\delta\theta= \delta r_1^a n_{2a}+r_1^a\delta n_{2a}=\delta r_1^a n_{2a}+\delta a_2\sin\theta.
\end{equation}
Thus we can decompose $\delta r_1^a$ as follows
\begin{equation}
    \delta r_1^a=(\cot\theta\delta \theta-\delta a_2)r_1^a+\dbar \tilde{B}_1^a
\end{equation}
with $\dbar \tilde{B}_1^a$ the tangential component of $\delta r_1^a$ to the corner. Furthermore, by $h_1^{ab}=r_1^ar_1^b+\gamma^{ab}$ at the corner, we can obtain the decomposition for $\delta h_1^{ab}$ in Eq. (\ref{ex0}). Likewise, one can obtain the decomposition for $\delta h_2^{ab}$. In addition, by $n_1^a n_{2a}=-\cos\theta$, we have
\begin{equation}
    \sin\theta\delta\theta=\delta n_1^an_{2a}+n_1^a\delta n_{2a}=\delta n_1^an_{2a}-\delta a_2\cos\theta=\dbar A_1^an_{2a}+(\delta a_1-\delta a_2)\cos\theta,
\end{equation}
which gives rise to the decomposition for $\dbar A_1^a$ in Eq. (\ref{ex0}) with $\dbar\tilde{A}_1^a$ the tangential component of $\dbar A_1^a$ to the corner. Similarly, one can obtain the decomposition for $\dbar A_2^a$.

Next by requiring $\delta g^{ab}|_{\Sigma_1}$
and  $\delta g^{ab}|_{\Sigma_2}$ in Eq. (\ref{ex1}) be equal to each other at the corner as well as contracting with $n_{1b}$ and $r_{1b}$, respectively, one will obtain
\begin{equation}
    \dbar\tilde{A}_1^a=-\dbar\tilde{A}_2^a\cos\theta+\dbar\tilde{B}_2^a\sin\theta,\quad \dbar\tilde{B}_1^a=\dbar\tilde{A}_2^a\sin\theta+\dbar\tilde{B}_2^a\cos\theta,
\end{equation}
which can be further cast into Eq. (\ref{re2}).

Finally, note that 
\begin{equation} \begin{aligned}
&\sum_{i=1,2}r_{ia} \omega_i^a =\sum_{i=1,2}(2r_{ia}\Psi^a{}_b\dbar{ A_i^b}+2r_{ia}h^{ae}\psi_{ecdb}n_i^d\delta h_i^{bc} )
=\sum_{i=1,2}(2r_{ia}\Psi^a{}_b\dbar{A_i^b}+2r_i^a\psi_{acdb}n_i^d\delta h_i^{bc})\\
=&\sum_{i=1,2}[2\psi_{acbd}r_i^an_i^cn_i^d\dbar{A_i^b}+2\psi_{acdb}r_i^an_i^d(r_i^b\dbar{\tilde B_i^c}+\delta \gamma^{bc})] \\
=& 2\psi_{acbd}r_1^an_1^cn_1^d\dbar{\tilde A_1^b}+2\psi_{acdb}r_1^an_1^dr_1^b( \cot\theta\dbar{\tilde A_1^c}+\csc\theta\dbar{\tilde A_2^c})+2\psi_{acdb}(r_1^an_1^d+r_2^an_2^d)\delta \gamma^{bc} \\
&+2\psi_{acbd}r_2^an_2^cn_2^d\dbar{\tilde A_2^b}+2\psi_{acdb}r_2^an_2^dr_2^b( \cot\theta\dbar{\tilde A_2^c}+\csc\theta\dbar{\tilde A_1^c})+\psi^{abcd}\varepsilon_{ab}\varepsilon_{cd}\delta\theta,
\end{aligned}\end{equation}
where $\psi_{acdb}=\psi_{[ac]db}$ has been used in the second line. Furthermore, the terms involving $\dbar{\tilde A_1^c}$ can be collected as follows
\begin{equation} \begin{aligned}
&( 2\psi_{abcd}r_1^an_1^bn_1^d-2\cot\theta\psi_{acbd}r_1^ar_1^bn_1^d+2\csc\theta\psi_{acdb}r_2^ar_2^bn_2^d)\dbar{\tilde A_1^c} \\
=& ( 2\psi_{abcd}r_1^an_1^bn_1^d-2\cot\theta\psi_{acbd}r_1^ar_1^bn_1^d+2\csc\theta\psi_{acbd}r_2^ar_1^bn_1^d)\dbar{\tilde A_1^c} \\
=& [ 2\psi_{abcd}r_1^an_1^bn_1^d+2\psi_{acbd}r_1^bn_1^d( r_2^a\csc\theta-r_1^a\cot\theta
 ) ]\dbar{\tilde A_1^c} \\
=& [ 2\psi_{abcd}r_1^an_1^bn_1^d+2\psi_{acbd}r_1^bn_1^d( n_1^a+r_1^a\cot\theta-r_1^a\cot\theta ) ]\dbar{\tilde A_1^c} \\
=& ( 2\psi_{abcd}r_1^an_1^bn_1^d-2\psi_{abcd}r_1^an_1^bn_1^d)\dbar{\tilde A_1^c} \\
=& 0,
\end{aligned}\end{equation}
where we have used $\psi_{a[bcd]}=0$ in the fifth line.
Similarly, one can show that the terms involving $\dbar{\tilde A_2^c}$ also cancel out. Therefore we end up with
\begin{equation} \begin{aligned}
\sum_{i=1,2}r_{ia} \omega_i^a =2\psi_{acdb}(r_1^an_1^d+r_2^an_2^d)\delta \gamma^{bc}+\psi^{abcd}\varepsilon_{ab}\varepsilon_{cd}\delta\theta,
\end{aligned}\end{equation}
which gives rise to Eq. (\ref{corner}).

\end{document}